\documentclass[nofootinbib,prd,12pt,superscriptaddress]{revtex4}%
\usepackage{amsmath}
\usepackage{amsfonts}
\usepackage{amssymb}
\usepackage{graphicx}%
\usepackage{amsmath,amssymb}
\usepackage{mathrsfs}
\usepackage{graphicx}
\usepackage{color}
\usepackage{subfigure}
\usepackage{fancyhdr}
\usepackage{multirow}
\usepackage{float}
\usepackage{epsfig}
\usepackage{amsfonts}
\usepackage{bm}

\def\be{\begin{equation}}
\def\ee{\end{equation}}
\def\ba{\begin{eqnarray}}
\def\ea{\end{eqnarray}}

\begin{document}

\title{Warm non-minimally coupled Peccei–Quinn Inflation \\and de Sitter Swampland Conjecture}

\author{Jureeporn Yuennan} 
\email{jureeporn_yue@nstru.ac.th}
\affiliation{Faculty of Science and Technology, Nakhon Si Thammarat Rajabhat University, Nakhon Si Thammarat, 80280, Thailand}

\author{Phongpichit Channuie}
\email{phongpichit.ch@mail.wu.ac.th}
\affiliation{School of Science, Walailak University, Nakhon Si Thammarat, 80160, Thailand}
\affiliation{College of Graduate Studies, Walailak University, Nakhon Si Thammarat, 80160, Thailand}

\author{Davood Momeni}
\affiliation{Northeast Community College, 801 E Benjamin Ave Norfolk, NE 68701, USA}

\date{\today}

\begin{abstract}
In this study, we explore the dynamics of warm inflation within a non-minimally coupled Peccei–Quinn (PQ) framework and evaluate its compatibility with the de Sitter Swampland Conjecture. Our model incorporates a PQ scalar field that is non-minimally coupled to gravity, facilitating inflation through a dissipative process that sustains a thermal bath, thereby distinguishing it from conventional cold inflation. We analyze the dissipation coefficient defined as $\Gamma(T, \sigma) = C_n T^n \sigma^p M^{1-n-p}$, where $C_n$ is a dimensionless constant, $M$ is a mass scale, and $n$ and $p$ are numerical powers. Our investigation focuses on three specific cases: (a) A temperature-dependent dissipation coefficient with an inverse relation, $\Gamma = C_{-1}\,\sigma^2/T$, where $n=-1$ and $p=2$; (b) A dissipation coefficient linear in field $\phi$, $\Gamma = C_{0} \sigma$, where $n=0$ and $p=1$; and (c) A dissipation coefficient linear in temperature $T$, $\Gamma = C_{1} T$, where $n=1$ and $p=0$. By examining the slow-roll dynamics in these inflationary scenarios, we derive essential cosmological parameters, including the scalar spectral index and the tensor-to-scalar ratio. We compare our results with the latest observational data from Planck 2018. Our findings suggest that the model is consistent with observational constraints while simultaneously satisfying the de Sitter Swampland conditions.
\end{abstract}

\maketitle

\section{Introduction}

Over the past decade, inflationary cosmology has become a cornerstone of modern theoretical physics, offering compelling explanations for the homogeneity, isotropy, and flatness of the universe. Inflation refers to a phase of rapid exponential expansion in the early universe, which effectively addresses key issues such as the horizon and flatness problems \cite{Starobinsky:1980te, Sato:1980yn, Guth:1980zm, Linde:1981mu, Albrecht:1982wi}. Traditional inflationary models are typically framed within the context of cold inflation (CI), where the universe cools as it expands, leading to the decoupling of the inflaton field from other fields. However, cold inflation presents challenges, particularly concerning the reheating phase, during which the universe must transition from the inflationary epoch to a hot, radiation-dominated era \cite{Linde:2005ht, Albrecht:1982mp, Abbott:1982hn}. This transition is crucial for explaining the observed cosmic microwave background (CMB) and the formation of light elements during Big Bang Nucleosynthesis (BBN).

In contrast, warm inflation (WI) offers an alternative scenario in which the inflaton field remains coupled to other fields throughout inflation, generating radiation alongside the expansion \cite{Berera:1995wh, Berera:1996fm, Berera:1999ws, Taylor:2000ze, Hall:2003zp, Berera:2008ar, Bartrum:2013fia}. This continuous energy dissipation into radiation allows the universe to maintain a thermal bath during inflation, thereby eliminating the need for a separate reheating phase. The dissipative dynamics of WI can significantly modify the inflationary trajectory, influencing key cosmological parameters such as the scalar spectral index, the tensor-to-scalar ratio, and the amplitude of primordial fluctuations. A potential realization of warm inflation driven by the self-interaction of the inflaton field was explored in Ref. \cite{Dymnikova:2000gnk}. Furthermore, various studies have examined both minimal and non-minimal couplings to gravity \cite{Panotopoulos:2015qwa, Benetti:2016jhf, Motaharfar:2018mni, Graef:2018ulg, Arya:2018sgw, Kamali:2018ylz}. More recently, Refs. \cite{Samart:2021eph, Samart:2021hgt, Cheng:2024uvn} have analyzed the Higgs-Starobinsky (HS) model and a non-minimally coupled scenario with a quantum-corrected, self-interacting potential within the warm inflation framework, see also Ref.\cite{Amaek:2021cqs} for warm inflation in general scalar-tensor theory of gravity. These developments render WI a compelling alternative to traditional cold inflation, particularly when considering recent observational constraints, such as those from Planck 2018 \cite{Planck:2018jri} and BICEP/Keck \cite{BICEP:2021xfz}.

The Peccei–Quinn (PQ) mechanism~\cite{Peccei:1977ur}, initially proposed to address the strong CP problem in quantum chromodynamics (QCD), introduces a PQ scalar field that spontaneously breaks a global U(1) symmetry, leading to the emergence of the axion~\cite{Weinberg:1977ma,Wilczek:1977pj}. In the context of inflation, this PQ scalar field can be coupled to gravity, resulting in a non-minimally coupled inflationary scenario. This combination of PQ symmetry breaking and non-minimal coupling provides a rich framework for exploring inflationary dynamics. Recent studies have focused on integrating PQ fields into warm inflationary models, enabling a unified description of both inflation and axion physics.

In this paper, we investigate a warm inflationary model featuring a non-minimally coupled PQ field, thereby extending the standard warm inflation scenario. We explore the dynamics of inflation within this framework and derive the slow-roll parameters, such as the scalar spectral index and tensor-to-scalar ratio, for comparison with observational data. Additionally, we examine the compatibility of this model with the de Sitter Swampland Conjecture, a theoretical criterion aimed at distinguishing effective field theories consistent with quantum gravity from those that reside in the so-called "Swampland." Subsequent studies \cite{Obied:2018sgi, Ooguri:2018wrx, Garg:2018reu} have suggested that the self-interaction potential of a scalar field governing the Universe's energy density must satisfy one of the following conditions:

\begin{equation}
\frac{|\nabla V|}{V} \geq \frac{c_{2}}{M_{p}}\,,
\end{equation}
or
\begin{equation}
\frac{{\rm min}(\nabla_{i}\nabla_{j} V)}{V} \leq -\frac{c_{3}}{M^{2}_{p}}\,.
\end{equation}

Here, $\nabla$ denotes the gradient in field space, while $c_2$ and $c_3$ are universal positive constants of order $1$. The term ${\rm min}(\nabla_i \nabla_j V)$ represents the minimum eigenvalue of the Hessian $\nabla_i \nabla_j V$ in an orthonormal frame. This implies that the potential must either be sufficiently steep or exhibit strong tachyonic instability. Ref. \cite{Kehagias:2018uem} argued that, given the current understanding regarding the origin of adiabatic curvature perturbations, slow-roll single-field inflation models remain compatible with the Swampland criteria and the current lower limit on the tensor-to-scalar ratio.

By analyzing the Peccei–Quinn warm inflation model under the Swampland framework, we aim to provide insights into the viability of warm inflation in addressing both theoretical and observational challenges. Our results demonstrate that the PQ warm inflation model remains consistent with observational bounds and satisfies the Swampland conditions, making it a promising candidate for describing early universe inflation. The structure of this work is as follows: we give a short review on cold \& warm non-minimal coupling scenario in Sec.\ref{Sec2}. Subsequently, we analyze the dissipation coefficient defined as $\Gamma(T, \sigma) = C_n T^n \sigma^p M^{1-n-p}$, where $C_n$ is a dimensionless constant, $M$ is a mass scale, and $n$ and $p$ are numerical powers. Our investigation focuses on three specific cases: (a) A temperature-dependent dissipation coefficient with an inverse relation, $\Gamma = C_{-1}\,\sigma^2/T$, where $n=-1$ and $p=2$; (b) A dissipation coefficient linear in field $\phi$, $\Gamma = C_{0} \sigma$, where $n=0$ and $p=1$; and (c) A dissipation coefficient linear in temperature $T$, $\Gamma = C_{1} T$, where $n=1$ and $p=0$. in Sec.\ref{Sec3}. In Sec.\ref{Sec4}, we compare the results in this work with the observational data and then in Sec.\ref{Sec5} examine if all cases satisfy the de Sitter swampland conjecture, with particular attention to entropy considerations. Our conclusions are drawn in the last section.

\section{Cold \& Warm Non-minimal coupling scenario}\label{Sec2}

\subsection{Review on Peccei-Quinn Inflation}
The Peccei–Quinn (PQ) mechanism~\cite{Peccei:1977ur}, which relies on a global \( U(1)_{PQ} \) symmetry, addresses the strong CP problem by promoting the QCD $\theta$-angle to a dynamic axion field~\cite{Weinberg:1977ma,Wilczek:1977pj}. The axion can be viewed as the phase of a complex scalar field, known as the PQ field. Once the PQ field settles at the minimum of a Mexican hat potential, the \( U(1)_{PQ} \) symmetry is spontaneously broken, and the axion emerges as a pseudo Nambu–Goldstone boson. Additionally, axions are considered a promising candidate for dark matter in the universe~\cite{Preskill:1982cy,Abbott:1982af,Dine:1982ah}.

We consider a PQ field $\Phi$ that is non-minimally coupled to the Ricci scalar in the Jordan frame \cite{Fairbairn:2014zta,DalCin:2023uai,Hamaguchi:2021mmt}:
\ba
 S_{J} = \int d^4 x \sqrt{-g}
\left[-
\left( \frac{M^2}{2} + \xi \Phi \Phi^* \right) R
+ g^{\mu \nu} \partial_\mu \Phi \partial_\nu \Phi^*
- V (\Phi) 
\right],
\ea
with $V$ being a Mexican hat potential,
\begin{equation}
 V (\Phi) = \frac{\lambda}{6} \left( |{\Phi}|^2 - \frac{f^2}{2}  \right)^2,
 \label{eq:Pqpotential}
\end{equation}
and $f$ is the axion decay constant.
The self-coupling constant~$\lambda$ and non-minimal coupling~$\xi$ are assumed to be non-negative. In general, $M$ is not automatically the Planck constant $M_{p}$. The non-minimal coupling to gravity is controlled by the dimensionless coupling $\xi$. Here the mass scale $M$ is related to the reduced Planck mass by
\begin{equation}
 M^{2}_{p} = M^{2}+2\xi \Phi \Phi^*\,.
 \label{eq:Pqpotential}
\end{equation}
At the symmetry-breaking vacuum, i.e., $\Phi=f/\sqrt{2}$, it becomes
\begin{equation}
 M^{2}_{p} = M^{2}+\xi f^{2}\,.
 \label{eq:Pqpotential1}
\end{equation}
Assuming that $\xi f^{2}\ll M^{2}_{p}$, we see that $M \simeq M_{p}$. Since $\Phi$ is the PQ complex scalar, we rewrite it in terms of its radial and
phase (axion) modes as
\begin{equation}
 \Phi = \frac{1}{\sqrt{2}}\phi e^{i\theta}\,.
 \label{eq:Pqpotential1}
\end{equation}
The angular component $\theta$ changes by a constant real value under the $U(1)_{PQ}$ rotation and corresponds to the QCD axion. Meanwhile, the radial component $\phi$ obtains a non-zero vacuum expectation value $\left<\phi\right>=f$, which leads to the spontaneous breaking of the PQ symmetry. The above action becomes
\ba
 S_{J} = \int d^4 x \sqrt{-g}
\left[
-\left( \frac{M^2+\xi \phi^{2}}{2}\right) R
+ \frac{1}{2}g^{\mu \nu} \partial_\mu \phi \partial_\nu \phi
+ \frac{1}{2}\phi^{2} g^{\mu \nu} \partial_\mu \theta \partial_\nu \theta- V(\phi) 
\right],
\ea
where 
\ba
V(\phi) =\frac{\lambda}{4!}\left(\phi^{2}-f^{2}\right)^{2}\,,
\ea
Note that for reducing isocurvature perturbations in this work we take $\phi\gg f$ during inflation. One mechanism by which this can be achieved is to take $\phi$ to be the inflaton. When transforming this action from the Jordan frame to the Einstein frame (which features a standard gravity sector), it is necessary to introduce a new scalar field to achieve canonical kinetic terms. We diagonalize the gravity-inflaton dynamics model via the conformal transformation: 
\ba
g_{\mu\nu}\rightarrow\tilde{g}_{\mu\nu}=\Omega({\phi})^2 g_{\mu\nu},\quad\Omega({\phi})^2=\frac{M^2+\xi\phi^{2}}{M_{p}^2},
\ea
such that 
\ba
\quad\tilde{g}^{\mu \nu}=\Omega^{-2}g^{\mu\nu},\quad\sqrt{-\tilde{g}}=\Omega^4\sqrt{-g}.
\ea
Applying the conformal transformation we land in the Einstein frame and the action reads:
\ba
S_{E} =\int d^{4}x \sqrt{-{\tilde {g}}}\left[-\frac{1}{2} M_{p}^2 \tilde {R} + \frac{1}{2} \tilde {g}^{\mu \nu} \partial_{\mu} \sigma \partial_{\nu} \sigma ++ \frac{1}{2}\frac{\phi^{2}}{\Omega^{2}} \tilde {g}^{\mu \nu} \partial_\mu \theta \partial_\nu \theta- U(\sigma(\phi))  \right]\,,
\ea
with
\ba
U(\sigma(\phi)) = \Omega^{-4}V(\phi).\label{potu}
\ea 
The $U(1)_{PQ}$ symmetry allows the axion to stay fixed to its initial position. Therefore, we treat the model as effectively single-field scenario. A canonically normalized field $\sigma$ related to $\phi$ via
\ba
\sigma'\equiv \frac{d\sigma}{d\phi}=\frac{\sqrt{1+(1+6 \xi)\frac{\xi \phi^2}{M_{p}^2}}}{1+\frac{\xi \phi ^2}{M_{p}^2}}\,.\label{sigp}
\ea
We will analyze the dynamics in the Einstein frame, and therefore define the slow-roll parameters in terms of $U$ and $\sigma$:
\ba
\epsilon = \frac{M_{p}^2}{2} \left( \frac{dU / d \sigma}{U} \right)^2, \quad \quad \eta = M_{p}^2 \left( \frac{d^2U / d \sigma^2}{U} \right), \quad \quad N = \frac{1}{M_{p}^2} \int _{\sigma_{end}} ^{\sigma_{ini}} \frac{U}{dU /d\sigma} d \sigma. \label{epsilon}
\ea
We will, however, express everything in terms of $\phi$, such that we don't need an explicit solution of (\ref{sigp}). We obtain:
\ba
\epsilon &=& \frac{M_{\rm P}^2}{2}  \left( -4 \Omega^{-1} \Omega' + \frac{V'}{V}\right) ^2 \left( \frac{1}{{\sigma}'}\right)^2 \nonumber\\&=& \frac{8 M_{p}^2}{\phi^2 \left(1+(6 \xi +1)\frac{\xi  \phi^2}{M_{p}^2}\right)}\,,\\\eta &=& \frac{M_{p}^2}{V\sigma^{'3}}\left(\frac{20 \sigma^{'} V \Omega^{'2}}{\Omega^2}-\sigma'' V'+\sigma' V''-\frac{8 \sigma' V' \Omega'}{\Omega}+\frac{4 V \sigma' \Omega'}{\Omega}-\frac{4 \sigma V \Omega''}{\Omega}\right)\nonumber\\&=&\frac{\frac{12 M_{p}^2}{\phi^2}+4 \xi  (12 \xi +1)-8 \xi  (6 \xi +1)\frac{\xi  \phi^2}{M_{p}^2}}{\left(1+(6 \xi +1)\frac{\xi  \phi^2}{M_{p}^2}\right)^2}\,,
\ea
and
\begin{align}
N = \frac{1}{M_{p}^2} \int _{\phi_{end}} ^{\phi_{ini}} \frac{V\sigma^{'2}}{-4 \Omega^{-1}\Omega' V + V'} d \phi =  \frac{1}{M_{\rm P}^2} \int _{\phi_{end}} ^{\phi_{ini}}\frac{\phi  \left(M_{p}^2+\xi  (6 \xi +1) \phi ^2\right)}{4 \left(M_{p}^2+\xi  \phi ^2\right)} d \phi \ . \label{efold}
\end{align}
We can roughly estimate the field value when inflation ends by solving $\epsilon=1$, giving,
\ba
\left(\frac{\phi_{\rm end}}{M_{p}^{2}}\right)^{2}\sim \frac{16}{\sqrt{(8 \xi +1) (24 \xi +1)}+1}\,,
\ea
which yields
\ba
\left(\frac{\phi_{\rm end}}{M_{p}^{2}}\right)^{2}\simeq 
\begin{cases}
8\quad\quad{\rm for}\quad \xi\ll 10^{-1}\,,\\
\frac{2}{\sqrt{3}\xi}\quad{\rm for}\quad \xi\gg 10^{-1}\,.
\end{cases}
\ea
\begin{figure}[ht!]
    \centering
\includegraphics[width=5in,height=5in,keepaspectratio=true]{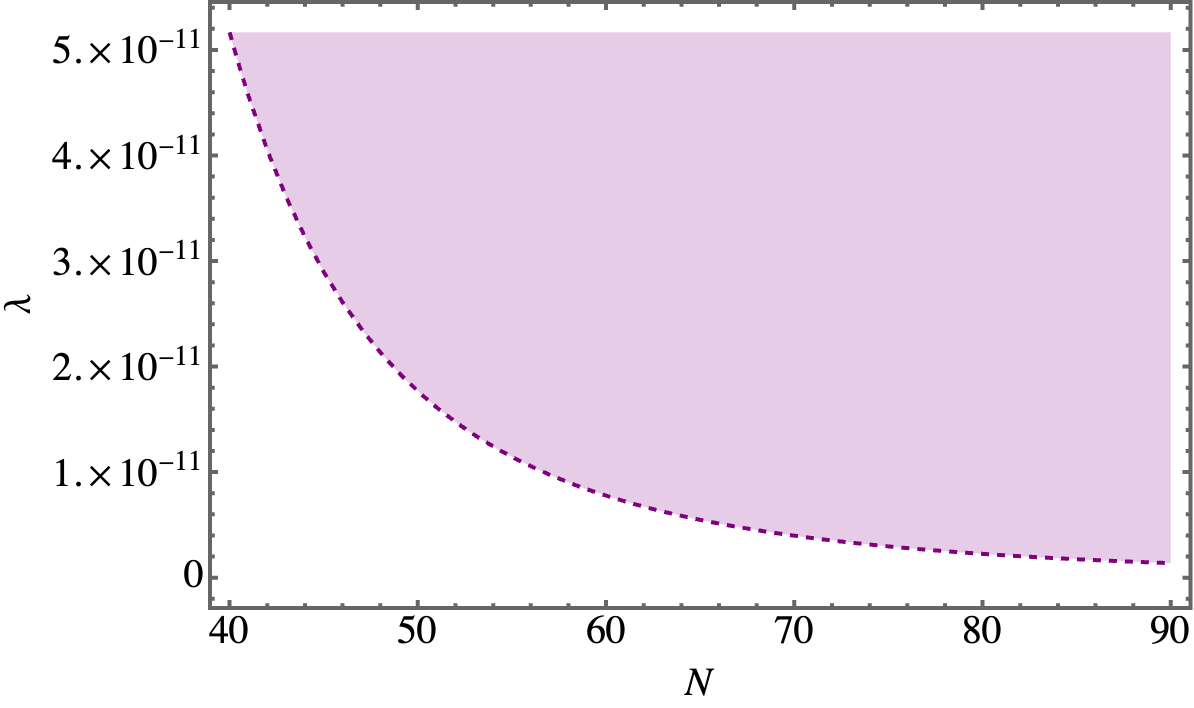}
    \caption{We display $\lambda$ as a function of e-foldings $N$ in which a shade follows a relation of Eq.(\ref{lamdaN111}).}
    \label{lambN}
\end{figure}
In Fig.(\ref{lambN}), we show the observational lower limit $\lambda$ as a function of e-foldings $N$ invoking the observational data.
Thus, as long as the decay constant $f$ meets the conditions $f\ll M_{\rm p}^{2}$ and satisfies equation (2.4), it implies that $f\ll \phi^{2}_{\rm end}$. This validates our decision to disregard $f$ during the inflationary period. In the large-field regime, we can show that the scalar power spectrum amplitude $A_{s}$, scalar spectral index $n_{s}$, and the tensor-to-scalar ratio $r$ can be written respectively as
\ba
A_{s}\simeq \frac{\lambda N^{2}}{72 \pi^{2}\xi (1+6\xi)}\,,\quad n_{s}\simeq 1-\frac{2}{N}\,,\quad r\simeq \frac{2(1+6\xi)}{\xi N^{2}}\,,
\ea
where we have assumed that $\phi^{2}_{N}\gg \phi^{2}_{\rm end}$. Using the scalar power spectrum amplitude takes the best-fit value $A_{s} = 2.1 \times 10^{-9}$, the coupling can be constrained as
\ba
\lambda \sim 1.49\times 10^{-6}\frac{\xi  (1+6\xi)}{N^2}\,.
\ea
Moreover, taking the latest BICEP/Keck data \cite{BICEP:2021xfz} for $r<0.036$, we find 
\ba
\xi >\frac{500}{9 N^2-3000}\,.
\ea
For instance, we find for $N=55\,(=60)$ that $\xi > 2.06\times 10^{-2}\,(>1.70\times 10^{-2})$, respectively. Using $\xi$ given above, we find for the lower limit of $\lambda$:
\ba
\lambda > \frac{8.29\times 10^{-5}}{\left(333.33 - N^2\right)^2}\,,\label{lamdaN111}
\ea
yielding for $N=55\,(=60)$ that $\xi > 1.14\times 10^{-11}\,(>7.77\times 10^{-12})$, respectively.

\subsection{Slow-roll dynamics in Peccei-Quinn Warm Inflation}
In a warm inflationary universe within the slow-roll regime, the Friedmann equation and the equations of motion for both the inflaton and the radiation matter can be reformulated as follows:
\begin{eqnarray}
H^2 &\approx& \frac{1}{3 M_p^2}\,U(\sigma)\,,
\label{SR-friedmann}
\\
\dot\sigma &\approx& -\frac{U'(\sigma)}{3H(1+Q)}\,,\qquad Q\equiv \frac{\Gamma}{3H}\,,
\label{SR-KG}
\\
\rho_r &\approx& \frac{\Gamma\,\dot\sigma}{4H}\,,\qquad \rho_r = C_r\,T^4\,,
\label{SR-rad}
\end{eqnarray}
where $Q$ denotes a dissipative coefficient and $C_r = g_*\,\pi^2/30$. To obtain the above expressions, we have used the following approximations:
\begin{eqnarray}
\rho_r &\ll& \rho_\sigma\,,\qquad 
\rho_\sigma = \frac12\,\dot\sigma^2 + U\,,
\\
\dot\sigma^2 &\ll& U(\sigma)\,,
\\
\ddot\sigma &\ll& 3H\left( 1 + Q\right)\dot\sigma\,,
\\
\dot\rho_r &\ll& 4H\,\rho_r\,,
\end{eqnarray}
As is typically done in the slow-roll scenario, warm inflation is modeled under the assumption that $Q\gg 1$ in the strong regime. Notably, the temperature can be expressed in terms of the scalar field, $\sigma$. In this study, we consider the temperature for $Q\gg 1$ to be represented in the following form:
\begin{eqnarray}
T &=& \left( \frac{U^{'2}\,\sigma^{m-1}}{4H\,C_m\,C_r}\right)^{\frac{1}{4+m}}\,,
\label{T-phi-strong}
\end{eqnarray}
with $m$ being any integer. Various choices for \(m\) have been explored in the literature \cite{Zhang:2009ge, Bastero-Gil:2011rva, Bastero-Gil:2012akf}. Specifically: (1) \(m = 1\), which corresponds to the high-temperature regime, as discussed in \cite{Berera:2008ar,Bastero-Gil:2016qru,Panotopoulos:2015qwa}; and (2) \(m = 3\), motivated by supersymmetric models \cite{Berera:2008ar,Bastero-Gil:2010dgy,Bastero-Gil:2011rva}, and featured in minimal warm inflation scenarios \cite{Berghaus:2019whh, Laine:2021ego, Motaharfar:2021egj}. In this work we consider three cases of $m$. In warm inflation, the slow-roll parameters are slightly modified and they take the form
\begin{eqnarray}
\epsilon &=& \frac{M_p^2}{2}\left( \frac{U'}{U}\right)^2=\frac{M_{\rm P}^2}{2}  \left( -4 \Omega^{-1} \Omega' + \frac{V'}{V}\right) ^2 \left( \frac{1}{{\sigma}'}\right)^2\,,\label{para1}\\\eta &=& M_p^2\,\frac{U''}{U}=\frac{M_{p}^2}{V\sigma^{'3}}\left(\frac{20 \sigma^{'} V \Omega^{'2}}{\Omega^2}-\sigma'' V'+\sigma' V''-\frac{8 \sigma' V' \Omega'}{\Omega}+\frac{4 V \sigma' \Omega'}{\Omega}-\frac{4 \sigma V \Omega''}{\Omega}\right)\,,\label{para2}\\\beta &=& M_p^2\left( \frac{U'\,\Gamma'}{U\,\Gamma}\right)=\frac{M_{p}^2 \Gamma'}{\Gamma  \sigma ^2}\left(\frac{V'}{V}-\frac{4 \Omega'}{\Omega' }\right)\,.
\label{para3}
\end{eqnarray}
Similar to cold inflation, inflationary phase of the universe in warm inflation takes place when the slow-roll parameters satisfy the following conditions:
\begin{eqnarray}
\epsilon \ll 1 + Q\,,\qquad \eta \ll 1 + Q\,,\qquad \beta \ll 1 + Q\,.\label{sloe}
\end{eqnarray}
Additionally, the number of e-foldings, $N$, in warm inflation gets modified and it can be written for $Q\gg 1$ as 
\begin{eqnarray}
N = \frac{1}{M_{p}^{2}}\int_{\sigma_{\rm end}}^{\sigma_N}\frac{Q\,U}{U'}\,d\sigma=\frac{1}{M_{p}^2} \int _{\phi_{end}} ^{\phi_{ini}} \frac{Q\,V\sigma^{'2}}{-4 \Omega^{-1}\Omega' V + V'} d \phi\,.\label{eqN}
\end{eqnarray}
The power spectrum of the warm inflation was calculated in Refs.\cite{Graham2009,Bastero-Gil:2018uep,Hall:2003zp,Ramos:2013nsa,BasteroGil:2009ec,Taylor:2000ze,DeOliveira:2001he,Visinelli:2016rhn}
and it reads
\begin{eqnarray}
\Delta_{\mathcal{R}}= \left( \frac{H_N^2}{2\pi\dot\sigma_N}\right)^2\left( 1 + 2n_N +\left(\frac{T_N}{H_N}\right)\frac{2\sqrt{3}\,\pi\,Q_N}{\sqrt{3+4\pi\,Q_N}}\right)G(Q_N)\,,
\label{spectrum}
\end{eqnarray}
Here, the subscript $``N"$ denotes quantities evaluated at the Hubble horizon crossing, and $n = 1/\big( \exp{H/T} - 1 \big)$ represents the Bose-Einstein distribution function. Additionally, the function $G(Q_N)$ captures the coupling between the inflaton and radiation within the heat bath, which results in a growing mode in the fluctuations of the inflaton field, as initially explored in Ref.\cite{Graham2009}, with subsequent implications \cite{BasteroGil:2009ec}. In addition, the scalar spectral index is defined as
\begin{eqnarray}
n_s - 1 = \frac{d\ln \Delta_{\mathcal{R}}}{d\ln k}\Bigg|_{k=k_N} = \frac{d\ln \Delta_{\mathcal{R}}}{dN}\,,
\end{eqnarray}
with $\ln k\equiv a\,H =N$\,. The tensor-to-scalar ratio of the perturbation, $r$, can be determined using the following formula:
\begin{eqnarray}
r = \frac{\Delta_T}{\Delta_{\mathcal{R}}}\,,
\label{tensor-scalar}
\end{eqnarray}
where $\Delta_T$ represents the power spectrum of the tensor perturbation and follows the same form as in standard (cold) inflation, i.e., $\Delta_T = 2H^2/\pi^2M_p^2 = 2U_{E}(\psi)/3\pi^2 M_p^4$. Thus, the power spectrum in Eq.(\ref{spectrum}) can be rewritten as follows:
\begin{eqnarray}
\Delta_{\mathcal{R}} = \frac{U_{E}(\phi_N)\big(1 + Q_N\big)^2}{24\,\pi^2\,M_p^4\,\varepsilon}\left(1 + 2\,n_N + \left(\frac{T_N}{H_N}\right)\frac{2\sqrt{3}\,\pi\,Q_N}{\sqrt{3+4\pi\,Q_N}}\right)G(Q_N)\,.
\end{eqnarray}
The growth rate of the inflaton field fluctuation, resulting from the coupling between the inflaton and the radiation fluid in the thermal bath, is described by the function $G(Q_N)$ \cite{Graham2009}. For Higgs-like and plateau-like potentials, the growing mode function was proposed in Ref.\cite{Bastero-Gil:2018uep} and is expressed as follows:
\begin{eqnarray}
G_1(Q_N) \simeq 1 + 0.18\,Q_N^{1.4} + 0.01\,Q_N^{1.8}\,,
\label{growing-mode}
\end{eqnarray}
whereas, the original growing mode function for warm little inflation was expressed as follows: \cite{Bastero-Gil:2016qru}
\begin{eqnarray}
G_2(Q_N) \simeq 1 + 0.335\,Q_N^{1.364} + 0.0185\,Q_N^{2.315}\,.
\label{growing-mode-original}
\end{eqnarray}
The function $G(Q)$ captures the growth of inflaton fluctuations due to coupling with radiation and must be basically determined. As noted in Ref.\cite{Bastero-Gil:2018uep}, this function also shows a slight dependence on the form of the scalar potential. For a quartic potential scenario, $G(Q)$ is provided in Eq.(\ref{growing-mode-original}), while for Higgs-like and plateau-like potentials, it is given in Eq.(\ref{growing-mode}). Note that at the thermalized inflaton fluctuation limit, $1+ 2\,n_N \simeq 2\,T_N/H_N $ and $T_N/H_N = 3\,Q_N/C_1$, one can re-write the power spectrum in the following form \cite{Bastero-Gil:2018uep},
\begin{eqnarray}
\Delta_{\mathcal{R}} \simeq \frac{5\,C_1^3}{12\,\pi^4\,g_*\,Q_N^2} \left(1+ \frac{\sqrt{3}\,\pi\,Q_N}{\sqrt{3+4\pi\,Q_N}}\right)G(Q_N)\,
\label{spectrum-grow},
\end{eqnarray}
where the approximation $\rho_r/V(\phi) = \varepsilon\,Q/2(1+Q)^2$ has been used to obtain above relation. We note that the above power spectrum in this limit is in-explicitly dependent on the inflaton potential \cite{Bastero-Gil:2018uep}. Then, the tensor-scalar ratio $r$ parameter in this case can be obtained by using Eqs.(\ref{tensor-scalar}) and (\ref{spectrum-grow}). It takes the form:
\begin{eqnarray}
r = \frac{\Delta_T}{\Delta_{\mathcal{R}}} = 16\,\varepsilon\left[ \frac{6\,Q_N^3}{C_1}\left(1+ \frac{\sqrt{3}\,\pi\,Q_N}{\sqrt{3+4\pi\,Q_N}}\right)G(Q_N)\right]^{-1}\,.
\label{tensor-scalar-growE}
\end{eqnarray}
The spectral index of the power spectrum, using the growing mode function in Eq.(\ref{growing-mode}), is given by \cite{Bastero-Gil:2018uep,BasteroGil:2009ec,Benetti:2016jhf}
\begin{eqnarray}
n_s &=& 1 + \frac{Q_N}{3+5\,Q_N}\frac{\big(6\,\epsilon - 2\,\eta \big)}{\Delta_{\mathcal{R}}}\,\frac{d\Delta_{\mathcal{R}}}{dQ_N}\,,
\label{ns-grow}\\
\frac{d\Delta_{\mathcal{R}}}{dQ_N} &=& \frac{5\,C_1^3}{12\,\pi^4\,g_*}\Bigg[\frac{1}{Q_N^2}\left(1+ \frac{\sqrt{3}\,\pi\,Q_N}{\sqrt{3+4\pi\,Q_N}}\right)\frac{dG(Q_N)}{dQ_N}
\nonumber\\
&-& \frac{1}{Q_N^3}\left(2+ \frac{\sqrt{3}\,\pi\,Q_N}{ \sqrt{3+4 \pi\,Q_N}}+\frac{2 \,\sqrt{3}\, \pi^2\,Q_N^2}{(3 + 4\,\pi\, Q_N)^{\frac32}}\right)G(Q_N) \Bigg].\nonumber
\label{diff-spectrum-growEns}
\end{eqnarray}
From Eq.(\ref{para1})-(\ref{para3}), we find
\ba
\epsilon(\phi)&=&\frac{8 M_{p}^2}{\phi^2 \left(1+(6 \xi +1)\frac{\xi  \phi^2}{M_{p}^2}\right)}\,,\\\eta(\phi)&=&\frac{\frac{12 M_{p}^2}{\phi^2}+4 \xi  (12 \xi +1)-8 \xi  (6 \xi +1)\frac{\xi  \phi^2}{M_{p}^2}}{\left(1+(6 \xi +1)\frac{\xi  \phi^2}{M_{p}^2}\right)^2}\,,\\\beta(\phi)&=&\frac{\frac{16 M_{p}^2}{\phi ^2}+48 \xi^2-16 \xi  (6 \xi +1)\frac{\xi  \phi^2}{M_{p}^2}}{5 \left(1+(6 \xi +1)\frac{\xi  \phi^2}{M_{p}^2}\right)^2}\,.
\ea

\section{Non-minimally coupled Peccei-Quinn Warm Inflation}\label{Sec3}
We take the growing mode function proposed in Ref.\cite{Bastero-Gil:2018uep} and it is expressed as follows:
\begin{eqnarray}
G(Q_N) \simeq 1 + 0.18\,Q_N^{1.4} + 0.01\,Q_N^{1.8}\,,
\label{growing-mode}
\end{eqnarray}
We only consider the slow-roll approximations in the large-field regime and $\xi\gg 10^{-1}$.
\subsection{$m=-1$ or $n=-1 \,\,\&\,\, p=2$}
Here in this case, we focus on $m=-1$ or $n=-1 \,\,\&\,\, p=2$ and then calculate $\Gamma(\phi)$ to obtain
\ba
\Gamma(\phi)&=&\frac{C_{-1} \sigma(\phi)^{'2}}{T(\phi)}\nonumber\\&=&\frac{2^{5/6} \sqrt[3]{3} C_{-1} \left(1+\frac{\xi  (1+6 \xi) \phi^2}{M_{p}^2}\right)}{\left(1+\frac{\xi  \phi^2}{M_{p}^2}\right)^2}\Bigg(\frac{\lambda  M_{p}^2 \phi^2}{C_{-1} C_{r} \left(\frac{\xi  (6 \xi +1) \phi^2}{M_{p}^2}+1\right)^2}\sqrt{\frac{\lambda  \phi^4}{M_{p}^2 \left(\frac{\xi  \phi^2}{M_{p}^2}+1\right)^2}}\Bigg)^{-1/3}\,.\label{ga}
\ea
Using Eq.(\ref{SR-KG}) and Eq.(\ref{ga}), we find $Q$:
\ba
Q=\frac{4 \sqrt[3]{6} \text{Cr} \text{Ct}^2 \text{Mp}^4 \left(\frac{\xi  (6 \xi +1) \phi ^2}{\text{Mp}^2}+1\right)^3}{\lambda ^2 \phi ^6}\left(\frac{\lambda  \phi ^2 \sqrt{\frac{\lambda  \phi ^4}{\text{Mp}^2 \left(\frac{\xi  \phi ^2}{\text{Mp}^2}+1\right)^2}}}{\text{Cr} \text{Ct} \left(\frac{\xi  (6 \xi +1) \phi ^2}{\text{Mp}^2}+1\right)^2}\right)^{2/3}\,.
\ea
When inflation ends, one finds using a condition $\epsilon_{\rm end}(\phi_{e}) \approx Q_{\rm end}(\phi_{e})$ that
\ba
\frac{8 M_{p}^2}{\phi_{e}^2 \left((6 \xi +1)\frac{\xi \phi_{e}^2}{M_{p}^2}\right)}\approx \frac{4 \sqrt[3]{6} C_{r} C_{-1}^2 \xi^3 (6 \xi +1)^3}{\lambda ^2 M_{p}^2}\left(\frac{\lambda M_{p}^4 \sqrt{\frac{\lambda M_{p}^2}{\xi ^2}}}{\text{Cr} C_{-1} \xi ^2 (1+6 \xi)^2 \phi ^2}\right)^{2/3}\,.\label{maineq}
\ea
Moreover, the inflaton field at the Hubble horizon crossing in the strong regime, $\phi_{N}$, can be determined using Eq.(\ref{eqN}) to yield
\begin{eqnarray}
N &=&\frac{1}{M_{p}^2} \int _{\phi_{end}} ^{\phi_{ini}} \frac{Q\,V\sigma^{'2}}{-4 \Omega^{-1}\Omega' V + V'} d \phi\nonumber\\&=&\int _{\phi_{end}} ^{\phi_{ini}}\frac{\sqrt[3]{6} C_{r} C_{-1}^2 M_{p}^2 \left(\frac{\xi  (6 \xi +1) \phi ^2}{M_{p}^2}+1\right)^{7/2}}{\lambda^2 \phi^5}\Bigg(\frac{\lambda  \phi ^2 \sqrt{\frac{\lambda  \phi ^4}{M_{p}^2 \left(\frac{\xi \phi^2}{M_{p}^2}+1\right)^2}}}{C_{r} C_{-1} \left(\frac{\xi  (6 \xi +1) \phi^2}{M_{p}^2}+1\right)^2}\Bigg)^{2/3}\,d \phi\,.\label{eqN}
\end{eqnarray}
However, the above equations can not be analytically solved to obtain exact solutions. Certain approximate solutions during inflation can be obtained by invoking a large field approximation. To begin with, we assume that $\xi\phi^{2}/M^{2}_{p}$ is much greater than 1. Subsequently, we divide the analysis into two scenarios: one where $\xi$ is much less than $10^{-1}$ and another where $\xi$ is much greater than $10^{-1}$. We again consider Eq.(\ref{maineq}) and solve for the second case of $\xi \gg 10^{-1}$ to obtain
\ba
\frac{\phi_{\rm e}}{M_{p}}=\frac{\lambda ^{3/8}}{\sqrt[8]{2} \sqrt{3} \sqrt[8]{C_{r}} \sqrt{C_{-1}} (\xi  (6 \xi +1))^{7/8}}\Big(\frac{(6 \xi +1)^2}{\xi^2}\Big)^{1/8}\,.
\ea
Moreover, the inflaton field at the Hubble horizon crossing in the strong regime, $\phi_{N}$, can be determined using Eq.(\ref{eqN}) for $\xi\gg 10^{-1}$ to yield
\ba
\frac{\phi_{N}}{M_{p}}=\frac{5^{3/5}}{\sqrt[5]{2} \,3^{4/5}}\Big(\frac{1}{C_{r} C_{-1} \xi^3 (6 \xi +1)^2}\Big)^{1/5}\Big(\frac{\lambda  \xi N}{C_{-1} (\xi  (6 \xi +1))^{3/2}}\Big)^{3/5}\,.
\ea
As done above, we therefore can re-write the slow-roll parameters in terms of the number of e-foldings, $N$, by using large field approximation in the strong Q limit and then we find. 
\ba
\epsilon_{N}&\simeq&\frac{216\ 2^{4/5} \sqrt[5]{3} C_{r} C_{-1} \xi^2 (1+6 \xi)}{25\, 5^{2/5} \left(\frac{\lambda  \xi  N}{C_{-1} (\xi  (1+6 \xi))^{3/2}}\right)^{12/5}}\Big(\frac{1}{C_{r} C_{-1} \xi ^3 (6 \xi +1)^2}\Big)^{1/5}\,,\\\eta_{N}&\simeq&-\frac{24\ 2^{2/5} 3^{3/5}}{5 \sqrt[5]{5} (6 \xi +1) \left(\frac{1}{C_{r} C_{-1} \xi ^3 (6 \xi +1)^2}\right)^{2/5} \left(\frac{\lambda  \xi  N}{C_{-1} (\xi  (6 \xi +1))^{3/2}}\right)^{6/5}}\,,\\\beta_{N}&\simeq&-\frac{48\ 2^{2/5} 3^{3/5}}{25 \sqrt[5]{5} (6 \xi +1) \left(\frac{1}{C_{r} C_{-1} \xi^3 (6 \xi +1)^2}\right)^{2/5} \left(\frac{\lambda  \xi  N}{C_{-1} (\xi  (6 \xi +1))^{3/2}}\right)^{6/5}}\,.
\ea
We can also write $Q(\phi=\phi_{N})$ as:
\ba
Q_{N}&\simeq& \frac{12\ 2^{3/5} 3^{2/5} C_{r} C_{-1}^2 \xi ^3 (6 \xi +1)^3}{5^{4/5} \lambda^2}\times \nonumber\\&&\times\Bigg(\frac{C_{-1}^2 (6 \xi +1)^3 \left(\frac{1}{C_{r} C_{-1} \xi ^3 (6 \xi +1)^2}\right)^{3/5} \left(\frac{\lambda  \xi Na}{C_{-1} (\xi  (6 \xi +1))^{3/2}}\right)^{4/5}}{N^2 \sqrt{\frac{\lambda }{\xi^2}}}\Bigg)^{2/3}\,.
\ea
In this section, we apply the COBE normalization condition \cite{Bezrukov:2008ut} to constrain the inflationary potentials. This condition allows us to determine the parameters of the models analyzed here. According to the Planck 2018 data, the inflaton potential must be normalized using the slow-roll parameter $\epsilon$ and must satisfy the following relation at the horizon crossing $\phi=\phi_{N}$ to match the observed amplitude of the cosmological density perturbations ($A_{s}$):
\begin{eqnarray}
\frac{U(\sigma(\phi)_N)}{\epsilon(\sigma(\phi)_N)} \simeq (0.0276\,M_{p})^{4}\,.
\label{Cobe-strong}
\end{eqnarray}
The constraints mentioned above are valuable for determining the model parameters, e.g., $\lambda,\,\xi$.
\ba
\lambda\simeq 7.12 \times 10^{-2}\Bigg(\frac{\xi^{5/17} }{(1+6 \xi)^{5/17} \left(\frac{1}{C_{r} C_{-1} \xi^3 (1+6 \xi)^2}\right)^{4/17} \left(\frac{\xi N}{C_{-1} (\xi  (1+6 \xi))^{3/2}}\right)^{12/17}}\Bigg)\,.
\ea
\begin{figure}[ht!]
    \centering
\includegraphics[width=5in,height=5in,keepaspectratio=true]{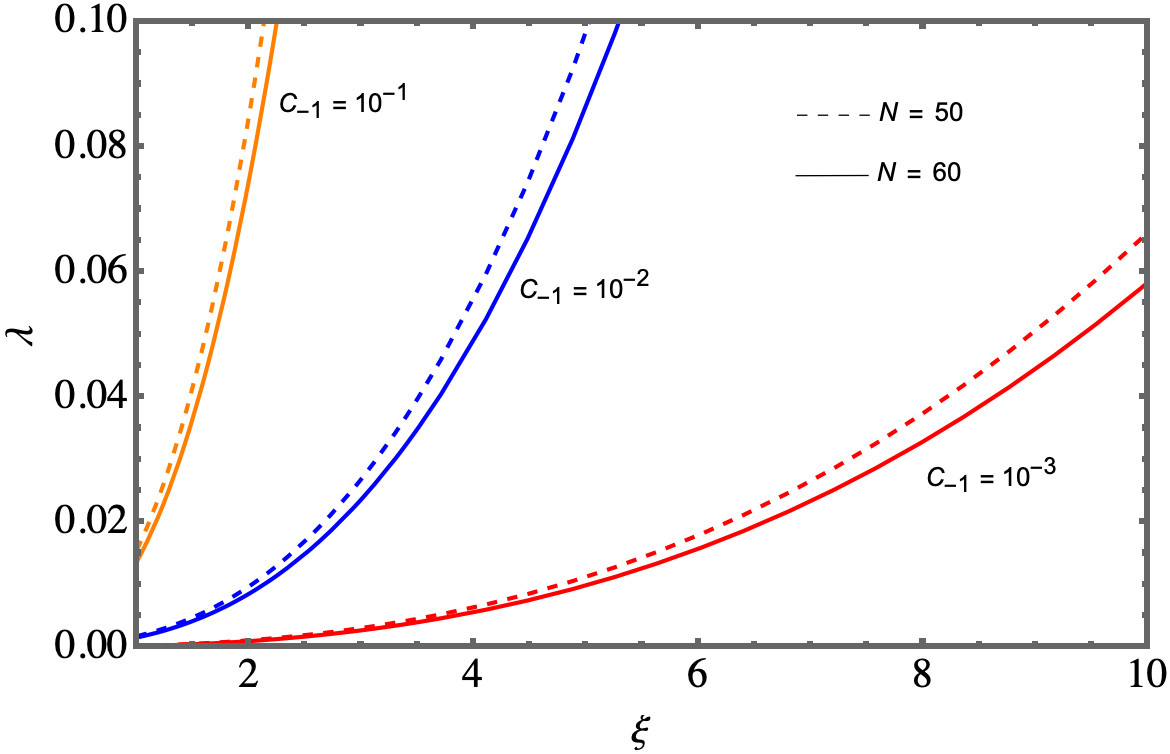}
    \caption{We present a plot of the parameter $\lambda$ as a function of the non-minimal coupling $\xi$, for different fixed values of the constants $C_{-1}$ and $N$.}
    \label{lambN11}
\end{figure}
The behavior of $\lambda$ as a function of $\xi$ is shown in Fig.~(\ref{lambN11}). It is evident that lower values of $C_{-1}$ lead to smaller values of $\lambda$.

\subsection{$m=0$ or $n=0 \,\,\&\,\, p=1$}
In the second case, we choose $m=0$ or $n=0 \,\,\&\,\, p=1$ and then calculate $\Gamma(\phi)$ to obtain
\ba
\Gamma(\phi)=C_{0}\, \phi(\sigma)=\frac{C_{0}}{1+\frac{\xi  \phi ^2}{M_{p}^2}}\sqrt{1+\frac{\xi  (1+6 \xi) \phi ^2}{M_{p}^2}}\,.\label{ga}
\ea
Using Eq.(\ref{SR-KG}) and Eq.(\ref{ga}), we find $Q$:
\ba
Q=\frac{2 C_{0} M_{p}}{\left(1+\frac{\xi  \phi^2}{M_{p}^2}\right) \sqrt{\frac{\lambda  \phi^4}{M_{p}^2 \left(\frac{\xi  \phi ^2}{M_{p}^2}+1\right)^2}}}\sqrt{1+\frac{2 \xi  (6 \xi +1) \phi ^2}{M_{p}^2}}\,.
\ea
When inflation ends, one finds using a condition $\epsilon_{\rm end}(\phi_{e}) \approx Q_{\rm end}(\phi_{e})$ that
\ba
\frac{8 M_{p}^2}{\phi_{e}^2 \left(1+(6 \xi +1)\frac{\xi \phi_{e}^2}{M_{p}^2}\right)}\approx\frac{2 C_{0} M_{p}}{\left(1+\frac{\xi  \phi_{\rm e}^2}{M_{p}^2}\right) \sqrt{\frac{\lambda  \phi_{\rm e}^4}{M_{p}^2 \left(\frac{\xi  \phi_{\rm e}^2}{M_{p}^2}+1\right)^2}}}\sqrt{1+\frac{2 \xi  (6 \xi +1) \phi_{\rm e}^2}{M_{p}^2}}\,.\label{maineq}
\ea
Moreover, the inflaton field at the Hubble horizon crossing in the strong regime, $\phi_{N}$, can be determined using Eq.(\ref{eqN}) to yield
\begin{eqnarray}
N &=&\frac{1}{M_{p}^2} \int _{\phi_{end}} ^{\phi_{ini}} \frac{Q\,V\sigma^{'2}}{-4 \Omega^{-1}\Omega' V + V'} d \phi\nonumber\\&=&\int _{\phi_{end}}^{\phi_{ini}}\frac{C_{0} \phi }{\sqrt{2} M_{p} \left(1+\frac{\xi  \phi^2}{M_{p}^2}\right) \sqrt{\frac{\lambda \phi^4}{M_{p}^2 \left(1+\frac{\xi \phi ^2}{M_{p}^2}\right)^2}}}\Big(1+\frac{\xi  (1+6 \xi) \phi ^2}{M_{p}^2}\Big)d\phi\,.\label{eqN}
\end{eqnarray}
However, the above equations can not be analytically solved to obtain exact solutions. Certain approximate solutions during inflation can be obtained by invoking a large field approximation. To begin with, we assume that $\xi\phi^{2}/M^{2}_{p}$ is much greater than 1. Subsequently, we divide the analysis into two scenarios: one where $\xi$ is much less than $10^{-1}$ and another where $\xi$ is much greater than $10^{-1}$. We again consider Eq.(\ref{maineq}) and solve for the second case of $\xi \gg 10^{-1}$ to obtain
\ba
\frac{\phi_{\rm e}}{M_{p}}=\frac{\sqrt[6]{2}}{\sqrt[3]{C_{0}} \sqrt[6]{54 \xi +9}}\Big(\frac{\lambda}{\xi^{5}}\Big)^{1/6}\,.
\ea
Moreover, the inflaton field at the Hubble horizon crossing in the strong regime, $\phi_{N}$, can be determined using Eq.(\ref{eqN}) for $\xi\gg 10^{-1}$ to yield
\ba
\frac{\phi_{N}}{M_{p}}=\frac{2 \sqrt{N}}{\sqrt{\sqrt{2} C_{0} (1+6 \xi)}}\Big(\frac{\lambda }{\xi^2}\Big)^{1/4}\,.
\ea
As done above, we therefore can re-write the slow-roll parameters in terms of the number of e-foldings, $N$, by using large field approximation in the strong Q limit and then we find. 
\ba
\epsilon_{N}\simeq\frac{(6 C_{0} \xi +C_{0})^2}{6 \lambda  N^2},\quad\eta_{N}\simeq-\frac{\sqrt{2} C_{0} (6 \xi +1)}{3 \xi  N \sqrt{\frac{\lambda }{\xi ^2}}},\quad\beta_{N}\simeq-\frac{4 \sqrt{2} C_{0}}{5 N \sqrt{\frac{\lambda }{\xi ^2}}}\,.
\ea
We can also write $Q(\phi=\phi_{N})$ as:
\ba
Q_{N}\simeq \frac{2^{3/4} \xi  (1+6 \xi) N}{\left(\frac{\xi  N \sqrt{\frac{\lambda }{\xi^2}}}{C_{0}}\right)^{3/2}}\,.
\ea
In this section, we apply the COBE normalization condition \cite{Bezrukov:2008ut} to constrain the inflationary potentials. This condition allows us to determine the parameters of the models analyzed here. According to the Planck 2018 data, the inflaton potential must be normalized using the slow-roll parameter $\epsilon$ and must satisfy the following relation at the horizon crossing $\phi=\phi_{N}$ to match the observed amplitude of the cosmological density perturbations ($A_{s}$):
\begin{eqnarray}
\frac{U(\sigma(\phi)_N)}{\epsilon(\sigma(\phi)_N)} \simeq (0.0276\,M_{p})^{4}\,.
\label{Cobe-strong}
\end{eqnarray}
The constraints mentioned above are valuable for determining the model parameters, e.g., $\lambda,\,\xi$.
\ba
\lambda\simeq \frac{0.00373185}{N}\Big(C_{0}^2 \xi^3 (1+6 \xi)\Big)^{1/2}\,.
\ea
\begin{figure}[ht!]
    \centering
\includegraphics[width=5in,height=5in,keepaspectratio=true]{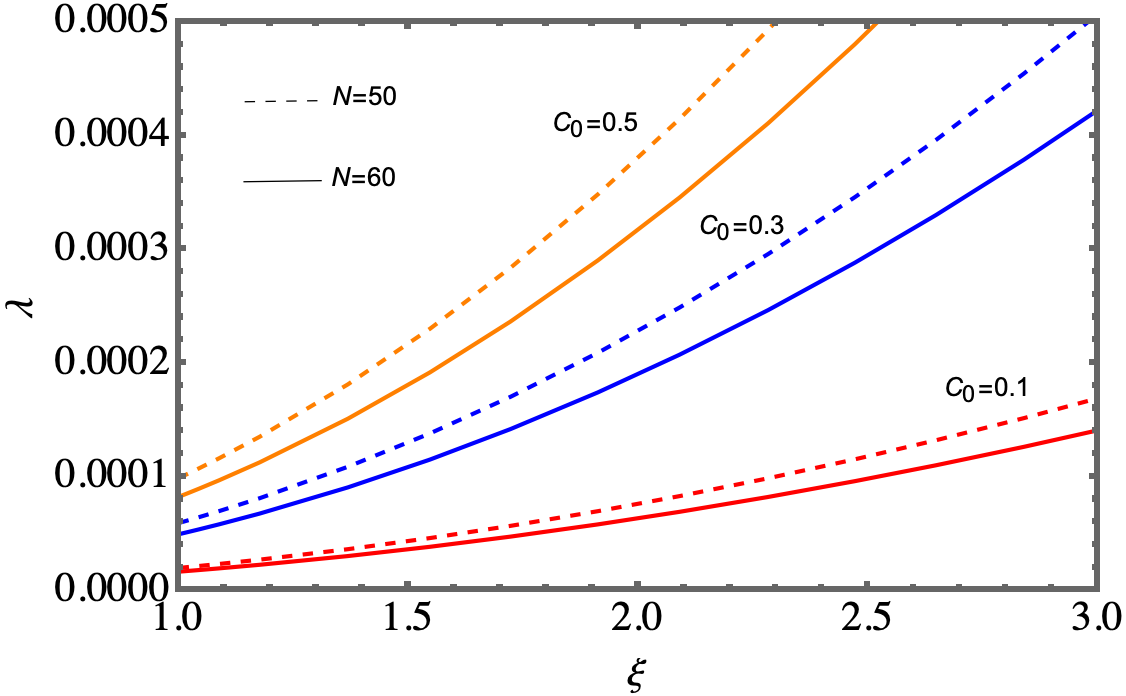}
    \caption{We present a plot of the parameter $\lambda$ as a function of the non-minimal coupling $\xi$, for different fixed values of the constants $C_{0}$ and $N$.}
    \label{lambN22}
\end{figure}
The behavior of $\lambda$ as a function of $\xi$ is shown in Fig.~(\ref{lambN22}). It is evident for any fixed value of $\xi$ that lower values of $C_0$ lead to smaller values of $\lambda$.
\subsection{$m=1$ or $n=1 \,\,\&\,\, p=0$}
In the last case, we consider $m=1$ or $n=-1 \,\,\&\,\, p=2$ and then calculate $\Gamma(\phi)$ to obtain
\ba
\Gamma(\phi)=C_{1} T_{p}(\phi)=\frac{1}{\sqrt{2} \sqrt[5]{3}}\frac{\left[C_{1}^4 \lambda^{3/2} M_{p} \phi^4 \left(1+\frac{\xi  \phi^2}{M_{p}^2}\right)^2\right]^{1/5}}{C_{r}\left[1+\frac{\xi  (6 \xi +1) \phi^2}{M_{p}^2}\right]^{1/5}}\,.\label{ga}
\ea
Using Eq.(\ref{SR-KG}) and Eq.(\ref{ga}), we find $Q$:
\ba
Q=\frac{2^{4/5}}{3^{2/5}}\left[\frac{\lambda  \phi^4}{M_{p}^2 \left(1+\frac{\xi  \phi^2}{M_{p}^2}\right)^2}\right]^{-1/2}\left[\frac{C_{1}^4 M_{p}^4 \left(\frac{\lambda \phi^4}{M_{p}^2 \left(1+\frac{\xi \phi^2}{M_{p}^2}\right)^2}\right)^{3/2}}{C_{r} \phi^2 \left(1+\frac{\xi  (1+6 \xi) \phi^2}{M_{p}^2}\right)}\right]^{1/5}\,.
\ea
When inflation ends, one finds using a condition $\epsilon_{\rm end}(\phi_{e}) \approx Q_{\rm end}(\phi_{e})$ that
\ba
\frac{8 M_{p}^2}{\phi_{e}^2 \left(1+(6 \xi +1)\frac{\xi \phi_{e}^2}{M_{p}^2}\right)}\approx\frac{2^{4/5}}{3^{2/5}}\left[\frac{\lambda  \phi\phi_{e}^4}{M_{p}^2 \left(1+\frac{\xi  \phi\phi_{e}^2}{M_{p}^2}\right)^2}\right]^{-1/2}\left[\frac{C_{1}^4 M_{p}^4 \left(\frac{\lambda \phi\phi_{e}^4}{M_{p}^2 \left(1+\frac{\xi \phi\phi_{e}^2}{M_{p}^2}\right)^2}\right)^{3/2}}{C_{r} \phi\phi_{e}^2 \left(1+\frac{\xi  (1+6 \xi) \phi\phi_{e}^2}{M_{p}^2}\right)}\right]^{1/5}\,.\label{maineq}
\ea
Moreover, the inflaton field at the Hubble horizon crossing in the strong regime, $\phi_{N}$, can be determined using Eq.(\ref{eqN}) to yield
\begin{eqnarray}
N &=&\frac{1}{M_{p}^2} \int _{\phi_{end}} ^{\phi_{ini}} \frac{Q\,V\sigma^{'2}}{-4 \Omega^{-1}\Omega' V + V'} d \phi\nonumber\\&=&\int _{\phi_{end}} ^{\phi_{ini}}\frac{C_{1} \phi }{2 \sqrt[5]{2}\, 3^{2/5} M_{p}^2}\left(\frac{\lambda  \phi ^4}{M_{p}^4 \left(1+\frac{\xi  \phi ^2}{M_{p}^2}\right)^2}\right)^{-1/2}\left(1+\frac{1+\xi  (6 \xi) \phi ^2}{M_{p}^2}\right)^{1/2}\nonumber\\&&\times\left[\frac{M_{p}^2 \left(\frac{\lambda  \phi ^4}{M_{p}^4 \left(\frac{\xi  \phi ^2}{M_{p}^2}+1\right)^2}\right)^{3/2}}{C_{1} C_{r} \phi ^2 \left(\frac{\xi  (6 \xi +1) \phi ^2}{M_{p}^2}+1\right)}\right]^{1/5}d \phi\,.\label{eqN}
\end{eqnarray}
However, the above equations can not be analytically solved to obtain exact solutions. Certain approximate solutions during inflation can be obtained by invoking a large field approximation. To begin with, we assume that $\xi\phi^{2}/M^{2}_{p}$ is much greater than 1. Subsequently, we divide the analysis into two scenarios: one where $\xi$ is much less than $10^{-1}$ and another where $\xi$ is much greater than $10^{-1}$. We again consider Eq.(\ref{maineq}) and solve for the second case of $\xi \gg 10^{-1}$ to obtain
\ba
\frac{\phi_{\rm e}}{M_{p}}=2^{11/16}\, 3^{1/8} \,\frac{C_{r}^{1/16} \lambda^{1/16}}{C_{1}^{1/4}} \left({\frac{1}{\xi^6 (1+6 \xi)^4}}\right)^{1/16}
\ea
Moreover, the inflaton field at the Hubble horizon crossing in the strong regime, $\phi_{N}$, can be determined using Eq.(\ref{eqN}) for $\xi\gg 10^{-1}$ to yield
\ba
\frac{\phi_{N}}{M_{p}}=\frac{3^{1/11}\, 22^{5/11}}{5^{5/11}}\frac{C_{r}^{1/11}}{C_{1}^{4/11} (\xi  (1+6 \xi))^{3/22}} N^{5/11} \left(\frac{\lambda}{\xi^{2}}\right)^{1/11}
\ea
As done above, we therefore can re-write the slow-roll parameters in terms of the number of e-foldings, $N$, by using large field approximation in the strong Q limit and then we find. 
\ba
\epsilon_{N}&\simeq&\frac{36\ 2^{4/5} \sqrt[5]{3}}{25\ 5^{2/5} \xi ^2 \left(\frac{1}{C_{r} C_{1} \xi^3 (6 \xi +1)^2}\right)^{4/5} \left(\frac{\lambda  \xi  N}{C_{1} (\xi  (6 \xi +1))^{3/2}}\right)^{12/5}}\,,\\\eta_{N}&\simeq&-\frac{4\ 2^{2/5} 3^{3/5}}{5 \sqrt[5]{5} \xi  \left(\frac{1}{C_{r} C_{1} \xi ^3 (6 \xi +1)^2}\right)^{2/5} \left(\frac{\lambda  \xi N}{C_{t} (\xi  (6 \xi +1))^{3/2}}\right)^{6/5}}\,,\\\beta_{N}&\simeq&-\frac{8 \sqrt[11]{\frac{3}{5}} 2^{4/11} C_{1}^{8/11} \left(\xi ^2\right)^{3/11}}{11^{10/11} C_{r}^{2/11} (6 \xi +1) N^{10/11} \left(\frac{\lambda }{\xi^2}\right)^{2/11}}\,.
\ea
We can also write $Q(\phi=\phi_{N})$ as:
\ba
Q_{N}\simeq \frac{2^{24/55}}{3^{26/55}}\left(\frac{5}{11}\right)^{4/11}\frac{C_{1}^{12/11}}{C_{r}^{3/11} (\xi  (1+6 \xi))^{1/11} N^{4/11}}\left(\frac{\lambda }{\xi ^2}\right)^{-3/11}\,.
\ea
In this section, we apply the COBE normalization condition \cite{Bezrukov:2008ut} to constrain the inflationary potentials. This condition allows us to determine the parameters of the models analyzed here. According to the Planck 2018 data, the inflaton potential must be normalized using the slow-roll parameter $\epsilon$ and must satisfy the following relation at the horizon crossing $\phi=\phi_{N}$ to match the observed amplitude of the cosmological density perturbations ($A_{s}$):
\begin{eqnarray}
\frac{U(\sigma(\phi)_N)}{\epsilon(\sigma(\phi)_N)} \simeq (0.0276\,M_{p})^{4}\,.
\label{Cobe-strong}
\end{eqnarray}
The constraints mentioned above are valuable for determining the model parameters, e.g., $\lambda,\,\xi$. 
\ba
\lambda\simeq 4.46\times 10^{-5} \left(\frac{C_{1}^{16/15} \xi^{14/15} (1+6 \xi)^{6/15}}{C_{r}^{4/15} N^{4/3}}\right)\,.
\ea
\begin{figure}[ht!]
    \centering
\includegraphics[width=5in,height=5in,keepaspectratio=true]{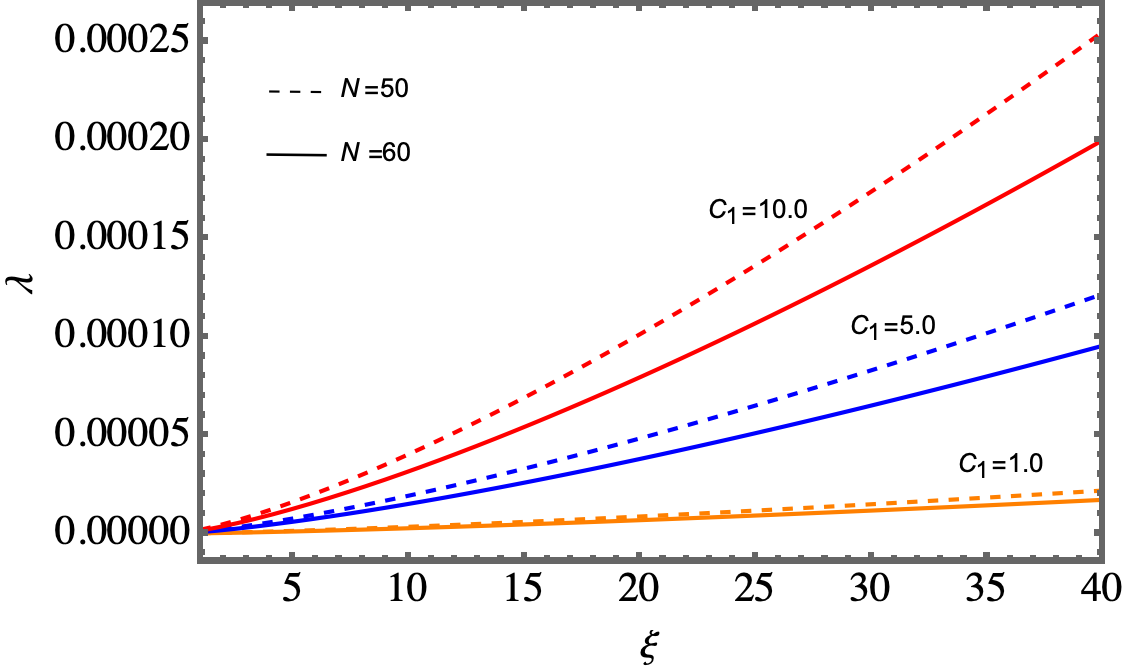}
    \caption{We present a plot of the parameter $\lambda$ as a function of the non-minimal coupling $\xi$, for different fixed values of the constants $C_{1}$ and $N$.}
    \label{lambN3}
\end{figure}
The behavior of $\lambda$ as a function of $\xi$ is shown in Fig.~(\ref{lambN3}). It is evident that lower values of $C_1$ lead to smaller values of $\lambda$.

\section{Confrontation with PLANCK2018 Data}\label{Sec4}

\subsection{$m=-1$ or $n=-1 \,\,\&\,\, p=2$}
\begin{figure}[ht!]
    \centering
\includegraphics[width=3.2in,height=3.2in,keepaspectratio=true]{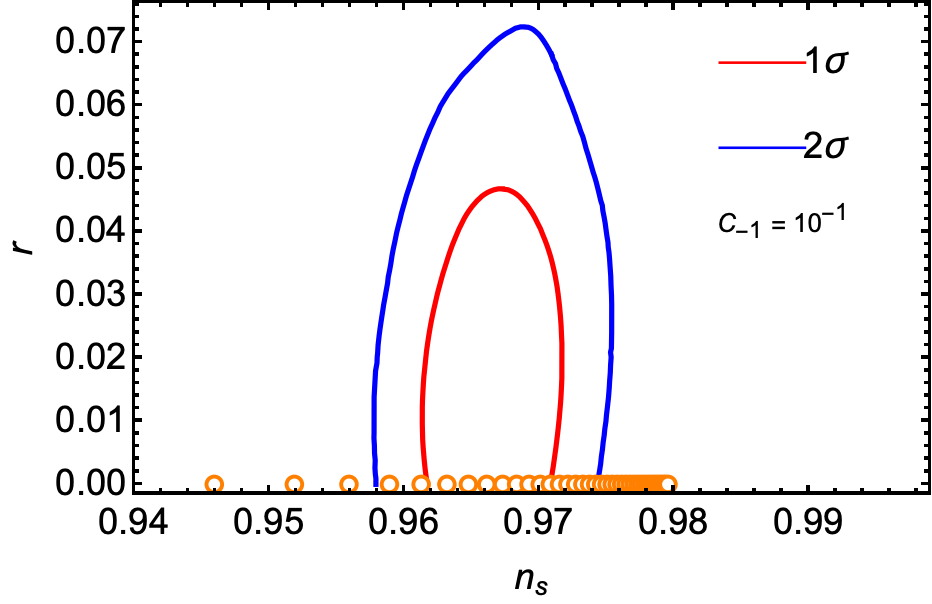}
\includegraphics[width=3.2in,height=3.2in,keepaspectratio=true]{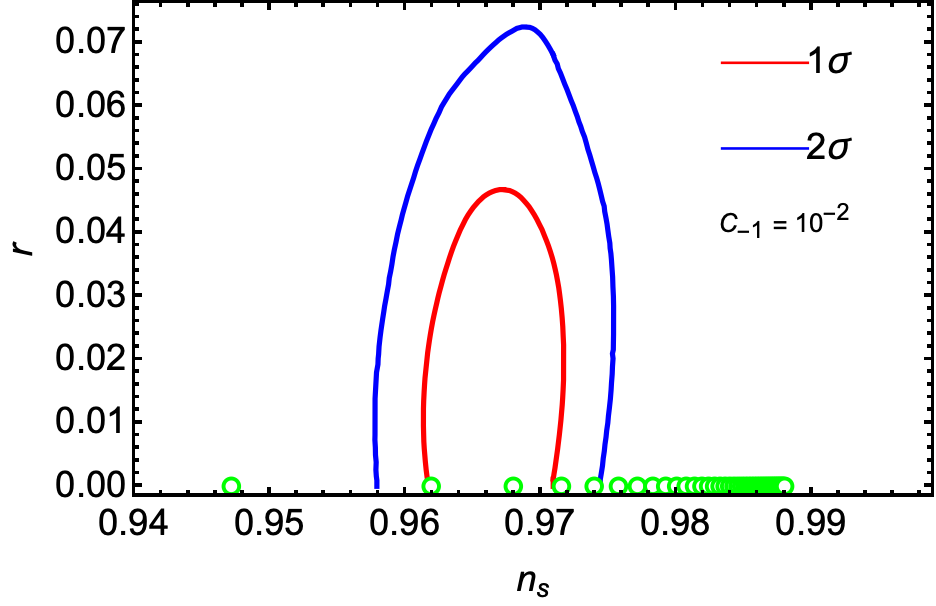}
\includegraphics[width=3.2in,height=3.2in,keepaspectratio=true]{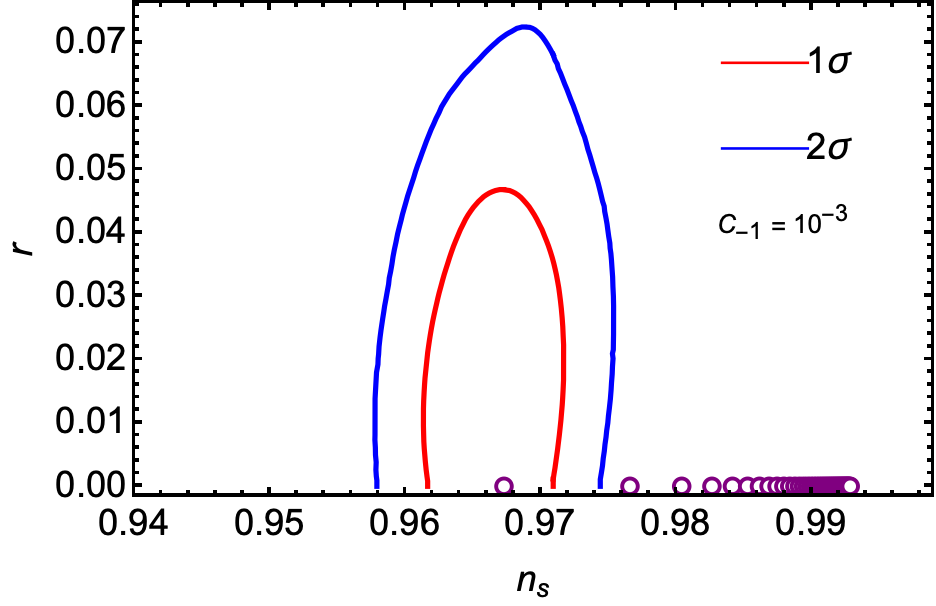}
    \caption{We compare the theoretical predictions of $(r,\,n_s)$ in the strong limit $Q\gg 1$ for $\Gamma = C_{-1}\,\phi^2/T$ ($n=-1$ and $p=2$) and $\xi\gg 10^{-1}$. We consider $G(Q)$ given in Eq.(\ref{growing-mode}) with $C_{r}=70$ and $N=60$. We consider theoretical predictions of $(r,\,n_s)$ for different values
of $C_{-1}=[10^{-1},10^{-2},10^{-3}]$ with Planck’18 results for TT, TE, EE, +lowE+lensing+BK15+BAO.}
    \label{lambN1}
\end{figure}

We plot the derived $n_{s}$ and $r$ for $G(Q)$ given in Eq.(\ref{growing-mode}) along with the observational constraints from Planck 2018 data in Fig.(\ref{lambN1}). In this case, we used three different values of $C_{-1}=[10^{-1},10^{-2},10^{-3}]$ and only considered $\xi\gg 10^{-1}$.  

We find for $C_{r}=70,\,C_{1}=10^{-1}$ and $N=60$ that $n_{s}=0.9647$ and $r=8.98\times 10^{-24}$ for $\xi\sim 10$, whilst $n_{s}=0.9680$ and $r=4.98\times 10^{-20}$ for $\xi\sim 3$. The model clearly predicts extremely small values of $r$. This model emphasizes a strong temperature dependence and predicts the lowest tensor-to-scalar ratio among the three cases, making it more challenging to detect tensor perturbations. This case might be suitable for scenarios requiring minimal gravitational waves, aligning with stringent observational bounds.

\subsection{$m=0$ or $n=0 \,\,\&\,\, p=1$}
\begin{figure}[ht!]
    \centering
\includegraphics[width=3.2in,height=3.2in,keepaspectratio=true]{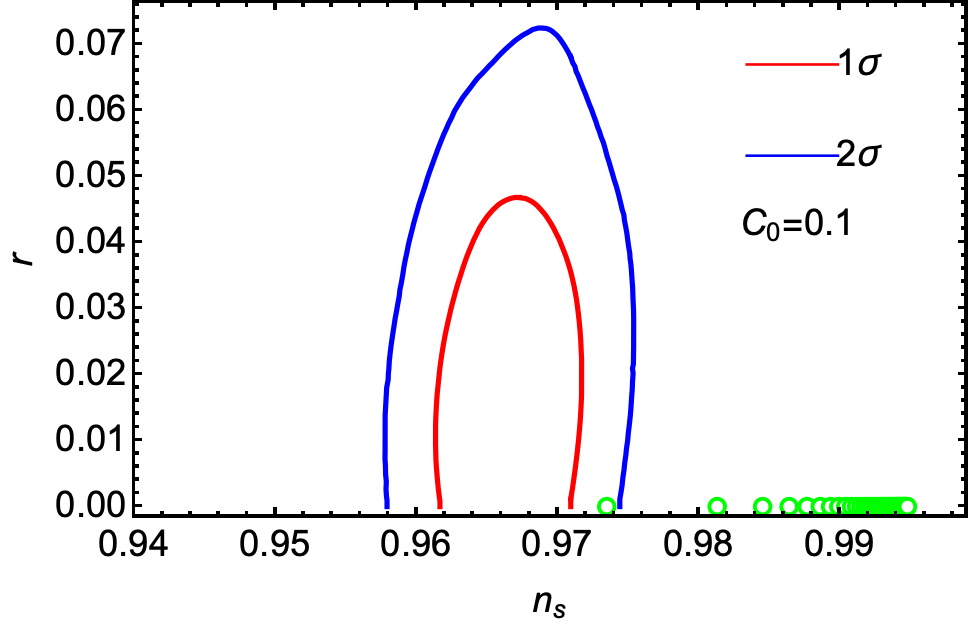}
\includegraphics[width=3.2in,height=3.2in,keepaspectratio=true]{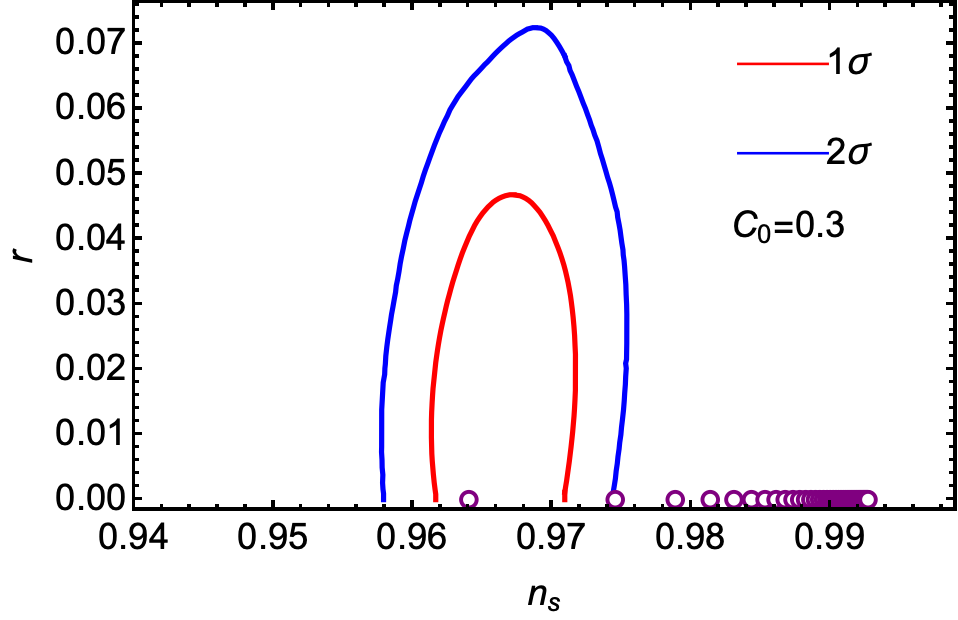}
\includegraphics[width=3.2in,height=3.2in,keepaspectratio=true]{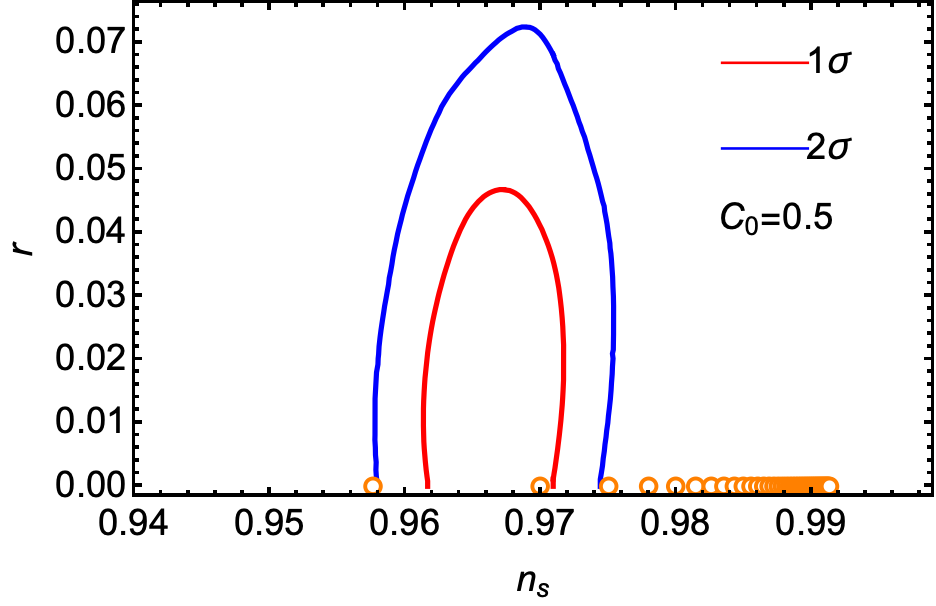}
    \caption{We compare the theoretical predictions of $(r,\,n_s)$ in the strong limit $Q\gg 1$ for $\Gamma = C_{0}\sigma(\phi)$ ($n=0$ and $p=1$) and $\xi\gg 10^{-1}$. We consider $G(Q)$ given in Eq.(\ref{growing-mode}) with $C_{r}=70$ and $N=60$. We consider theoretical predictions of $(r,\,n_s)$ for different values
of $C_{-1}=[0.1,0.3,0.5]$ with Planck’18 results for TT, TE, EE, +lowE+lensing+BK15+BAO.}
    \label{lambN2}
\end{figure}

We display the derived $n_{s}$ and $r$ for $G(Q)$ given in Eq.(\ref{growing-mode}) along with the observational constraints from Planck 2018 data in Fig.(\ref{lambN2}). In this case, we used three different values of $C_{0}=[0.1,0.3,0.5]$ and also considered $\xi\gg 10^{-1}$.  

We discover for $C_{r}=70,\,C_{1}=0.5$ and $N=60$ that $n_{s}=0.9699$ and $r=7.50\times 10^{-14}$ for $\xi\sim 2$, whilst $n_{s}=0.9640$ and $r=9.54\times 10^{-13}$ for $\xi\sim 1$. The model also predicts extremely small values of $r$. The linear field dependence of $\Gamma$ leads to slightly larger values of $r$ compared to the $m=-1$ case. This model represents a compromise between minimal gravitational waves and a robust fit to the spectral index.

\subsection{$m=1$ or $n=1 \,\,\&\,\, p=0$}
We display the derived $n_{s}$ and $r$ for $G(Q)$ given in Eq.(\ref{growing-mode}) along with the observational constraints from Planck 2018 data in Fig.(\ref{lambN3}). In this case, we used three different values of $C_{0}=[1,5,10]$ and also considered $\xi\gg 10^{-1}$.
\begin{figure}[ht!]
    \centering
\includegraphics[width=3.2in,height=3.2in,keepaspectratio=true]{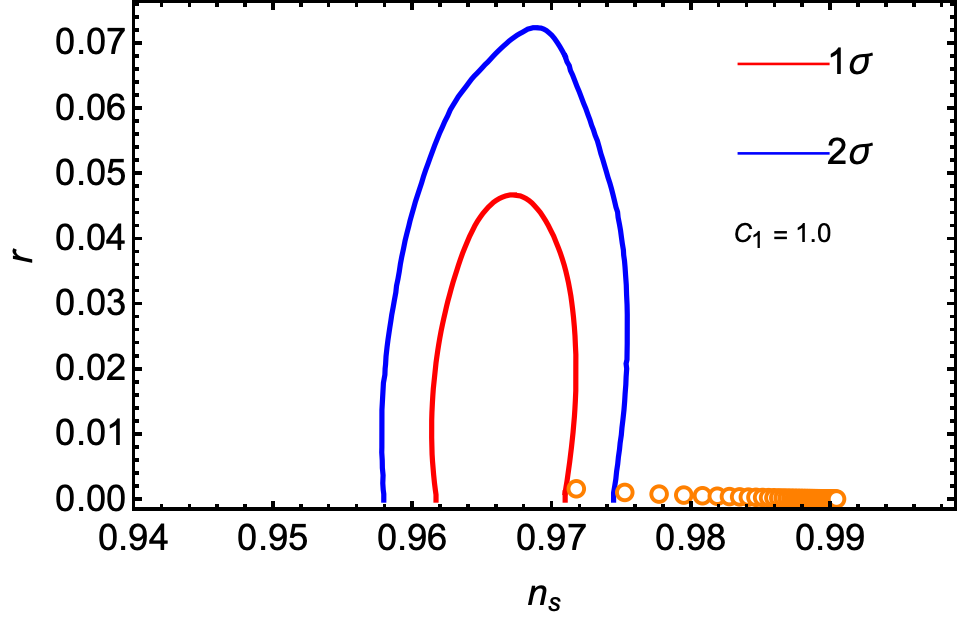}
\includegraphics[width=3.2in,height=3.2in,keepaspectratio=true]{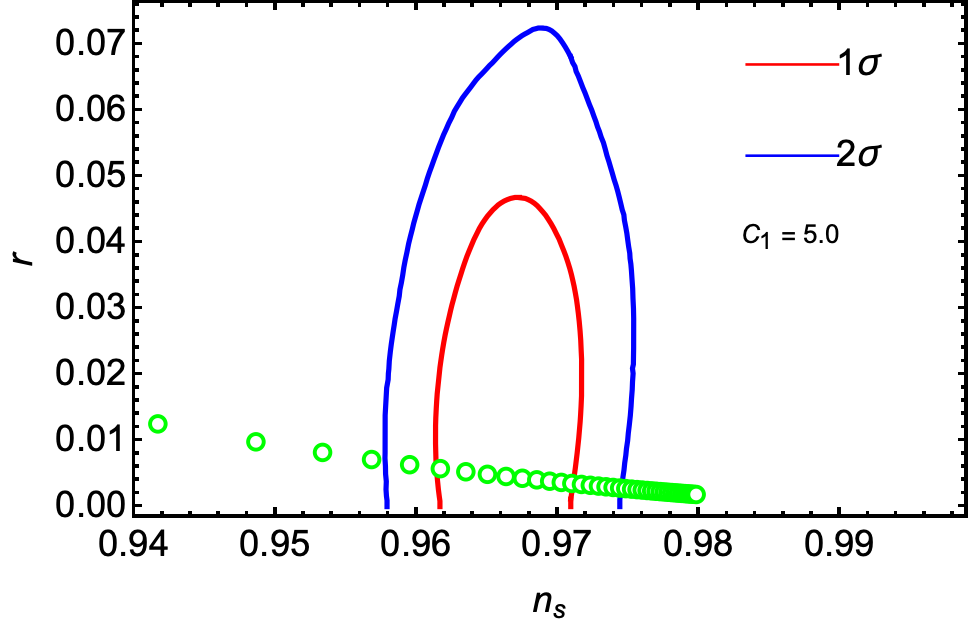}
\includegraphics[width=3.2in,height=3.2in,keepaspectratio=true]{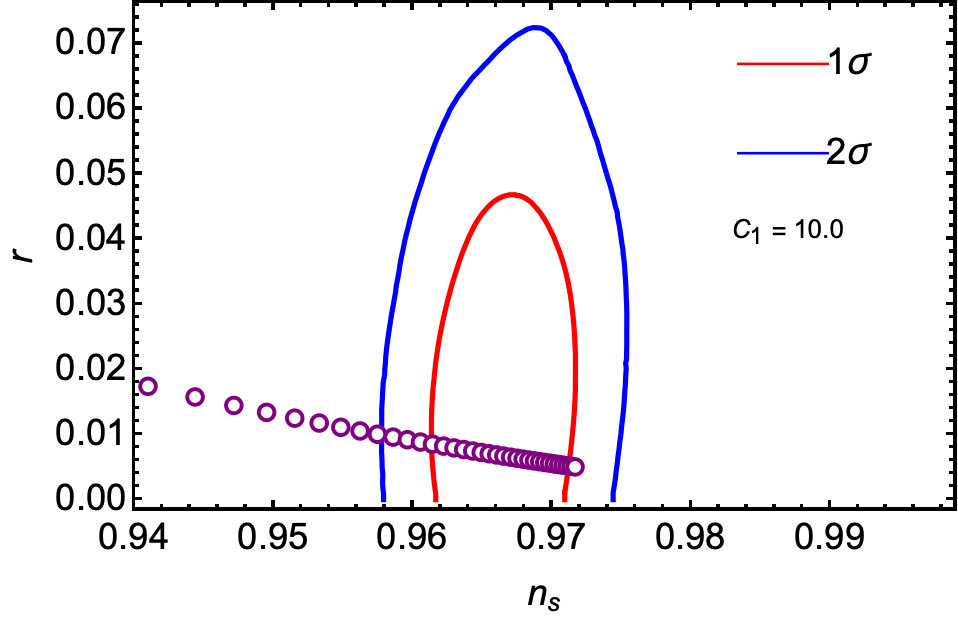}
    \caption{We compare the theoretical predictions of $(r,\,n_s)$ in the strong limit $Q\gg 1$ for $\Gamma = C_{0}\sigma(\phi)$ ($n=1$ and $p=0$) and $\xi\gg 10^{-1}$. We consider $G(Q)$ given in Eq.(\ref{growing-mode}) with $C_{r}=70$ and $N=60$. We consider theoretical predictions of $(r,\,n_s)$ for different values
of $C_{1}=[1,5,10]$ with Planck’18 results for TT, TE, EE, +lowE+lensing+BK15+BAO.}
    \label{lambN3}
\end{figure}

We find for $C_{r}=70,\,C_{1}=5$ and $N=60$ that $n_{s}=0.9650$ and $r=0.0049$ for $\xi\sim 9$, whilst for $C_{r}=70,\,C_{1}=10$ $n_{s}=0.9649$ and $r=0.0073$ for $\xi\sim 23$. The model also predicts extremely small values of $r$. This model exhibits the strongest dependence on temperature, leading to the highest values of $r$. It is more optimistic for detecting gravitational waves, though still at the lower observational limit.

\section{Warm Inflation and de Sitter swampland conjecture}\label{Sec5}
\begin{figure}[ht!]
    \centering
\includegraphics[width=5in,height=5in,keepaspectratio=true]{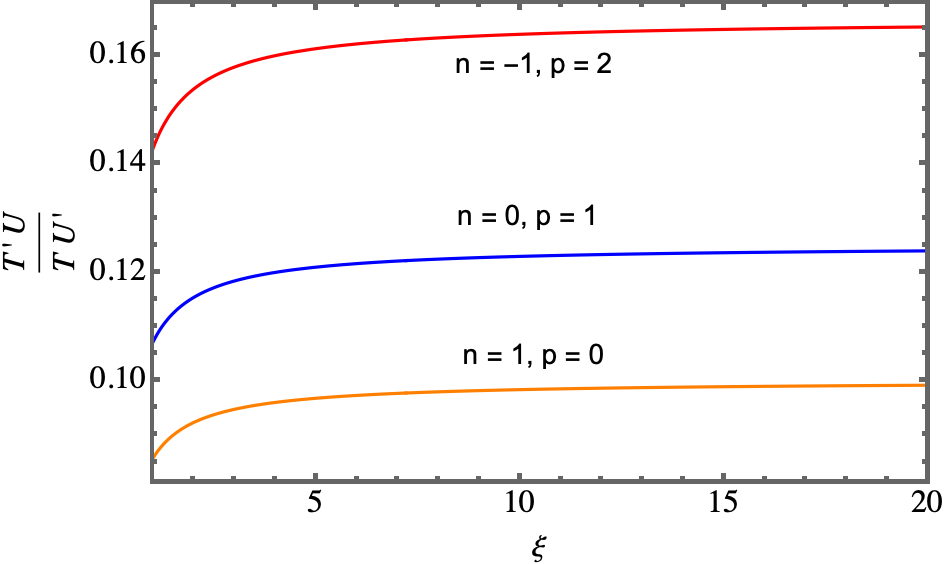}
    \caption{A plot of the ratio $\frac{T' U}{T U'}\big|_{Q\gg 1}$ versus $\xi$ for $n=-1$ and $p=2$, $n=0$ (red) and $p=1$, and $n=1$ (blue) and $p=0$ (orange).}
    \label{TUUT}
\end{figure}

In this section, we closely follow previous work of which the authors demonstrated the robustness of warm inflation (WI) in addressing the de Sitter swampland conjectures, with particular attention to entropy considerations \cite{Brandenberger:2020oav}. In WI framework, the inflaton interacts with other fields, generating dissipation and radiation throughout inflation. Unlike Cold Inflation (CI), where the Universe cools down as it expands, WI allows for continuous radiation production, maintaining a thermal bath during inflation. This changes the inflaton dynamics, which are described by
\begin{equation}
\ddot{\phi} + (3H + \Gamma)\dot{\phi} + U' = 0,
\end{equation}
where $\Gamma$ is the dissipation coefficient, and $H$ is the Hubble parameter. The energy density of the radiation fluid evolves according to
\begin{equation}
\dot{\rho_r} + 4H\rho_r = \Gamma \dot{\phi}^2,
\end{equation}
with $\rho_r$ being the radiation energy density. The entropy density $s$ is related to the radiation energy by $Ts = (1 + w_r)\rho_r$, where $w_r = 1/3$ for a thermalized radiation bath, and $\rho_r = \alpha T^4$ with $\alpha = \pi^2 g_*/30$ and $g_*$ as the number of degrees of freedom. During WI, the key slow-roll equations become
\begin{equation}
3H(1+Q)\dot{\phi} \approx -U', \quad \rho_r \approx \frac{3Q}{4}\dot{\phi}^2, \quad Ts \approx Q\dot{\phi}^2,
\end{equation}
where $Q = \Gamma / 3H$ denotes the dissipation rate. Since the vacuum energy dominates over radiation ($U \gg \rho_r$), the Hubble parameter is approximately
\begin{equation}
H^2 \approx \frac{U}{3M_{Pl}^2}.
\end{equation}
Combining these expressions leads to an important relation for the temperature
\begin{equation}
\alpha T^4 \approx \frac{(U')^2}{12H^2} \frac{Q}{(1+Q)^2}.\label{tem}
\end{equation}
Now, considering the Bousso entropy bound, where the total entropy inside a Hubble volume must not exceed the Gibbons-Hawking entropy, we modify the condition to account for the thermal bath's entropy. This gives
\begin{equation}
N\gamma (M_{Pl}R)^\delta + \frac{4}{3}\pi sR^3 \leq 8\pi^2 (M_{Pl}R)^2.
\end{equation}
This leads to a modified upper bound on the number of degrees of freedom
\begin{equation}
N\gamma M_{Pl}^\delta \leq R^{3-\delta}\left( 8\pi^2 R^{-1}M_{Pl}^2 - \frac{4}{3} \pi s \right),
\end{equation}
where $R = 1/H$. Using the Friedmann equation in the slow-roll approximation, and differentiating with respect to the inflaton field $\phi$, we obtain
\begin{equation}
\frac{|U'|}{U} > c_2 + \delta c_2 \equiv \tilde{c}_2,
\end{equation}
where $c_2$ is universal and positive constants of order 1$, \delta c_2$ is a small correction due to the entropy from radiation and  $\tilde{c}_2$ given by
\begin{eqnarray}
\tilde{c}_2 =  c_2\left| 1-\frac{1}{3-\delta}
\frac{1-\frac{\dot{s}}{6\pi M_{\rm Pl}^2\dot{H}}} {1+\frac{s}{6\pi
    M_{\rm Pl}^2 H}}\right|^{-1}.
\label{c2tilde}
\end{eqnarray}  
To estimate the WI modification to the de Sitter swampland conjecture, we calculate the entropy terms. First, we find the ratio of entropy density to the Hubble parameter
\begin{equation}
\frac{s}{H} \approx \frac{Q}{1+Q} \frac{2M_{p}^2}{H} \frac{\epsilon_H}{T} \quad \text{where} \quad \epsilon_H\equiv \frac{\epsilon_U}{1+Q},\, \epsilon_U \approx \frac{M_{Pl}^2}{2} \left( \frac{U'}{U} \right)^2.
\end{equation}
Since $T > H$ in WI, the ratio $s/(6\pi M_{Pl}^2 H)$ is much smaller than 1, which implies
\begin{equation}
\frac{s}{6\pi M_{Pl}^2 H} \ll 1.
\end{equation}
This result holds in both weak ($Q \ll 1$) and strong ($Q \gg 1$) dissipation regimes, and remains valid throughout inflation, even as $\epsilon_H \to 1$ at the end of inflation. Next, we consider the term $\dot{s}/(6\pi M_{Pl}^2 \dot{H})$, which can be expressed as
\begin{equation}
\frac{\dot{s}}{6\pi M_{Pl}^2 \dot{H}} = \frac{s}{6\pi M_{Pl}^2 H} \frac{s'H}{sH'},
\end{equation}
where $s'H/sH' \approx 6 T'V / (T V')$. Substituting this into $\tilde{c}_2$, hence, Eq.~(\ref{c2tilde}) becomes
\begin{eqnarray}
\tilde{c}_2 \simeq  c_2  \Big| 1 -
\frac{1}{3-\delta} \left[ 1+ \kappa \left( 1 - 6 \frac{T' V}{T V'}
  \right) +  {\cal O }( \kappa^2 )
  \right]\Big|^{-1},
\label{c2tilde2}
\end{eqnarray}  
where
\begin{equation}
\kappa = \frac{s}{6 \pi M_{\rm Pl}^2 H} .
\end{equation}
To examine the coefficient $\tilde{c}_2$ in WI, we parametrize the dissipation coefficient $\Gamma$ as \cite{Berera:2008ar,Bartrum:2013fia,Bastero-Gil:2010dgy,Bastero-Gil:2012akf,Zhang:2009ge}
\begin{eqnarray}
\Gamma(T,\phi) = C T^n \phi^p M^{1-n-p},\label{gama}
\end{eqnarray}
where $C$ is a dimensionless constant, $M$ is a mass scale, and $n$ and $p$ are numerical powers. This form of $\Gamma$ covers various WI models. Stability of WI requires $-4 < n < 4$ in both weak ($Q \ll 1$) and strong ($Q > 1$) regimes. From Eqs.~(\ref{tem}) and (\ref{gama}), in the high dissipative regime ($Q > 1$), we get
\begin{eqnarray}
T(\phi) \simeq \left[ \frac{\sqrt{3} M_{\rm Pl}}{4 \alpha} \frac{(U')^2}{C \phi^p U^{1/2} M^{1-n-p}} \right]^{\frac{1}{4+n}},
\end{eqnarray}
and
\begin{eqnarray}
\frac{T' U}{T U'}\Bigr|_{Q>1} \simeq  \frac{1}{4+n} \left[ 2\frac{U U''}{(U')^2} - \frac{p U}{\phi U'} - \frac{1}{2} \right].\label{test}
\end{eqnarray}
Using the results given by Eq.(\ref{test}), we can quantify whether our model of warm inflation can be made compatible with the swampland conditions. We will check with the inflaton potential in the present work. Let us consider the case of inflation models parametrized by the the primordial potential (\ref{potu}). This form of the potential is known to satisfy the observational constraints in the cold inflation scenario. Using Eq.(\ref{potu}) in Eq.(\ref{test}), we find
\begin{eqnarray}\label{UT}
\frac{T' U}{T U'}\bigg|_{Q\gg 1} \sim \begin{cases}
\frac{3 \xi }{5 (6 \xi +1)}\quad{\rm for}\quad n=1,\,p=0,\\
\frac{3 \xi }{4 (6 \xi +1)}\quad{\rm for}\quad n=0,\,p=1,\\
\frac{\xi }{6 \xi +1}\quad\quad{\rm for}\quad n=-1,\,p=2.
\end{cases}
\end{eqnarray}
We display the behaviors of $\frac{T' U}{T U'}\big|_{Q\gg 1}$ dependent on $\xi$ in Fig.(\ref{TUUT}).
In the equations above, we consider three well-known dissipation coefficients in warm inflation (WI) with a quartic inflaton potential (\ref{potu}): (a) A temperature-dependent dissipation coefficient with an inverse relation~\cite{Bartrum:2013fia}, $\Gamma = C_{-1}\,\phi^2/T$, where $n=-1$ and $p=2$. This leads to a value of $\frac{T' U}{T U'}\big|_{Q\gg 1}\sim \frac{3 \xi }{5 (6 \xi +1)}$ for Eq.(\ref{UT}); (b) A dissipation coefficient linear in field $\phi$, $\Gamma = C_{0} \phi(\sigma)$, where $n=0$ and $p=1$. This gives $\frac{T' U}{T U'}\big|_{Q\gg 1}\sim \frac{3 \xi }{4 (6 \xi +1)}$ for Eq.(\ref{UT}); and (c) A dissipation coefficient linear in temperature $T$, $\Gamma = C_{1} T$, where $n=1$ and $p=0$. This gives $\frac{T' U}{T U'}\big|_{Q\gg 1}\sim \frac{\xi }{6 \xi +1}$ for Eq.(\ref{UT}). In all cases, the ratio $\frac{T' U}{T U'}\big|_{Q\gg 1}$ is generally less than 1, see Fig.(\ref{TUUT}), resulting in $\tilde{c}_2 \simeq c_2$ for these warm non-minimally-coupled PQ inflation models.

\section{Conclusion}
In conclusion, this study investigates the dynamics of warm inflation within a non-minimally coupled Peccei–Quinn (PQ) framework, integrating elements of the de Sitter Swampland Conjecture. We developed a warm inflation model based on a PQ field, which demonstrated compatibility with current observational data, such as the Planck 2018 results. The model was analyzed in both cold and warm inflationary scenarios, where the inflaton was shown to dissipate energy into radiation, maintaining a thermal bath throughout inflation. This dissipation altered the inflaton dynamics, distinguishing the warm inflation scenario from traditional cold inflation.

Our findings suggest that warm inflation is robust in light of the Swampland conjectures, particularly due to its capability to generate a consistent thermal bath during inflation. The slow-roll dynamics of the inflaton and radiation were examined, with detailed calculations showing that our model can fit within observational bounds. The results also indicate that the dissipative coefficient plays a crucial role in modifying the traditional slow-roll parameters. All models produce spectral indices $n_{s}$ close to the observed range ($n\approx 0.964$), indicating robust theoretical consistency. However, they differ in how tightly they fit the constraints on $r$. The $m=1$ model is attractive for future observational efforts due to its higher predicted $r$ reaching detectable levels by future experiments like LiteBIRD \cite{LiteBIRD:2022cnt} or CMB-S4 \cite{Abazajian:2019eic}.

Lastly, our exploration of the de Sitter Swampland Conjecture revealed that the entropy contributions from the thermal bath remain within acceptable limits, confirming that the model satisfies the conjecture's conditions. Thus, the Peccei–Quinn warm inflation scenario not only fits within the current cosmological constraints but also aligns with theoretical predictions related to the Swampland conjecture, making it a compelling candidate for describing early universe inflationary dynamics.

\begin{acknowledgments}
This work is financially supported by Thailand NSRF via PMU-B under grant number PCB37G6600138.
\end{acknowledgments}

\end{document}